# Ferroelectric domain scaling and electronic properties in ultrathin BiFeO₃ films on vicinal substrates

Vilas Shelke,[1*] Dipanjan Mazumdar,[1] Sergey Faleev[1], Oleg Mryasov[1,2], Stephen Jesse,[3] Sergei Kalinin,[3] Arthur Baddorf,[3] and Arunava Gupta[1]

[1]Center for Materials for Information Technology, University of Alabama, Tuscaloosa, AL, 35487 USA

[2]Department of Physics, University of Alabama, Tuscaloosa, AL, 35487 USA

[3]Center for Nanophase Materials Sciences, Oak Ridge National Laboratory, Oak Ridge, TN 37831 USA

## Abstract

We report electrically switchable polarization and ferroelectric domain scaling over a thickness range of 5-100 nm in BiFeO₃ films deposited on [110] vicinal substrates. The BiFeO₃ films of variable thickness were deposited with SrRuO₃ bottom layer using pulsed laser deposition technique. These films have fractal domain patterns and the domain width scales closely with the square root of film thickness, in accordance with the Landau-Lifschitz-Kittel (LLK) law. The Switching Spectroscopy Piezo-response Force Microscopy provides clear evidence for the ferroelectric switching behavior in all the films. Using Quasi-particle Self-consistent GW (QPGW) approximation we have investigated physical parameters relevant for direct tunneling behavior, namely the effective mass and effective barrier height of electrons. For rhombohedral BFO, we report a large effective barrier height value of 3.6 eV, which is in reasonable agreement with nanoscale transport measurements. QPGW investigations into the tetragonal BFO structure with *P4mm* symmetry revealed a barrier height of 0.38 eV, significantly lower compared to its rhombohedral counterpart. This difference has very significant implications on the transport properties of nearly tetragonal BFO phase.

* Corresponding author, present address- Department of Physics, Barkatullah University, Bhopal- 462 026 INDIA; e mail- drshelke@gmail.com

# I. INTRODUCTION

In the era of miniaturization and power conservation each quantum of power and size matters a lot. The particularly influenced are the information storage devices. The topical trend is the spin polarized tunneling with ferroelectric control, which relies on zero current gate controlled operation with low power dissipation.[1] The use of electrically switchable ferroelectric tunnel junctions for nonvolatile memory devices[2,3] is providing an impetus for extensive study on ultrathin ferroelectric films[4-7]. In this context, $BiFeO_3$ is a material with huge potential mainly due to it's multiferroic character, which provides ability to switch effectively by electric or magnetic field.[8] Intriguingly, it may also turn out to be power saving material with fundamentally different photovoltaic effect.[9,10] Therefore, deposition and observation of ferroelectricity in ultrathin $BiFeO_3$ films is of current research interest.

The recent studies on $BiFeO_3$ (BFO) thin films are guided by two aspects viz. tetragonality and vicinality. The BFO films deposited on large lattice mismatched substrates like $LaAlO_3$ show monoclinically distorted tetragonal structure.[11,12] The stabilization of pure tetragonal phase remains elusive although the theoretical calculations suggest strain[13] and electric field[14] induced phase stabilization or transition. We have reported ferroelectric behavior of strain relaxed $BiFeO_3$ thin films on lattice mismatched substrates[15] and have also pursued the tetragonality issue for the fundamental understanding using Piezo-response Force Microscopy (PFM) and polarized Raman Spectroscopy techniques.[16,17] On the other hand, deposition of BFO films on vicinal substrates can assist engineering of ferroelectric domain structure.[18,19] Such films have two domain variants as compared to four domain variants observed in thin films deposited on plain substrates. As a consequence the ferroelectric switching is better on vicinal substrates. Our study on BFO films deposited on $SrTiO_3$ substrates having $4^0$ miscut along [110] direction revealed higher polarization value and significantly reduced coercive field.[20] It is clear that the domain structure plays a vital role in governing the technologically important parameters like polarization and coercive field.

In BFO films, the ferroelectric domains are large in dimensions and are stripe or fractal types.[8] Generally, the domains follow the Landau, Lifshitz and Kittel (LLK) law wherein the domain width scales with the square root of film thickness.[21] However, deviation from this law is recently reported in very thin BFO films deposited on plain $SrTiO_3$ substrates.[22] Therefore, from fundamental and application point of view, it is pertinent to study the ferroelectric behavior and domain structure in ultrathin films deposited on vicinal substrates. For the present work, we deposited 5-100 nm thickness BFO films on $SrTiO_3$ substrates with $4^0$ miscut along [110] direction. We confirmed the structural quality of the films using X-ray diffraction technique. The fractal domains with variable dimensions were observed using Piezo-response Force Microscopy. We tried to fit the domain size variation with the film thickness in accordance with LLK law. We also provide unambiguous evidence for the ferroelectricity in ultra-thin (up to 5 nm) BFO films using advanced Switching Spectroscopy Piezo-response Force Microscopy (SSPFM) technique. For spintronics application as ferroelectric barrier in a magnetic tunnel junction estimating the potential barrier height for tunneling is the quantity of interest. To that effect, we have conducted theoretical and experimental investigations into the electronic transport properties of BFO. Using Quasi-particle Self-consistent GW technique and local current-voltage measurements we find a reasonable agreement in the barrier height values for the R*3c* symmetry BFO. On the other hand, similar band structure analysis for the tetragonal *P4mm* symmetry gave a much lower effective barrier height value compared to rhombohedral BFO, making this phase extremely promising for device applications.

## II. EXPERIMENTAL

We have deposited $BiFeO_3$ (BFO) thin films of variable thickness in the range 5-100 nm on (100) $SrTiO_3$ substrates with $4^0$ miscut along [110] direction. The BFO films with $SrRuO_3$ (SRO) bottom layer were deposited sequentially using pulsed laser deposition method.[23] The crystalline quality and out of plane lattice parameters were determined using X-ray Diffraction (X'pert Pro. Panalytical) with $CuK_\alpha$ radiation. The surface topography and domain structure were revealed using Atomic and Piezo-response Force Microscopy under ambient conditions. For this purpose, commercial Scanning Probe Microscope (Cypher, Asylum Research) equipped with Pt coated conducting tip (AC240TM, Olympus) was operated at the resonance frequency around 260 kHz and ac bias amplitude of 2V. The same set up was augmented to perform advanced Switching Spectroscopy Piezo-response Force Microscopy (SS-PFM) measurements.[24,25] We used 25 x 25 grid on 2 μm x 2 μm scan area to map the local polarization switching with variable tip bias between +10 V and -10 V. Similarly, we measured local current vs voltage (I-V) characteristics at intermediate voltage of ± 2 V. Details about the *ab-initio* methods are described later in an appropriate section.

## III. RESULTS AND DISCUSSION

### A. Structural Analysis

The X-ray diffraction patterns (θ-2θ scan) of 100, 50, 20 and 10 nm BFO films deposited with 40 nm SRO bottom layer on STO substrates with $4^0$ miscut along [110] direction are shown in Fig. 1 (a). The inset shows clearly distinguishable (100) peaks. The highly oriented textured growth of the films is evident from the appearance of *(00l)* peaks.

The representative omega and phi scans shown in Fig. 1 (b) and (c) clearly reveal the crystalline quality and epitaxial nature of these films. The full width half maxima (FWHM) values for the omega scan were around 0.04 $^0$ and the phi scan exhibited four–fold symmetry. The out of plane lattice parameters determined from XRD data were 4.01, 4.05, 4.06, and 4.08 Å for the 100, 50, 20 and 10 nm thickness films. The bulk value for the psudo-cubic lattice parameter is 3.96 Å. However, the substrate induced in-plane compression of unit cells causes out of plane elongation. A variation in lattice parameter has been reported with the substrates[11,26], with film thickness[27] and also with operating oxygen pressure[28]. Our films seem somewhat constrained with reducing thickness, which is in agreement with the reported behavior for other miscut substrates.[18] On plain substrates, the change in lattice parameters for the BFO films below 100 nm thickness is quite small.[5,27]

## B. Ferroelectric domain scaling and switching behavior

The surface topography scans of 2 μm x 2 μm area of the 100, 50, 20, 10 and 5 nm thickness BFO films are shown in Fig. 2 (a)-(e) respectively. In general BFO films are known to grow in 3-D island growth mode.[18] However, step flow growth mode has been reported on orthorhombic DyScO$_3$ substrates[29], on the miscut substrates[18,26] and possibly with growth under optimized oxygen pressure[28]. The parallel steps with larger width resulting from step bunching were reported on the substrates with $4^0$ miscut along [100] direction.[18,26] On the other hand, when miscut angle is along [110] direction, as in the present case, the steps have saw-tooth pattern.[20] The parallel steps promote nanowire kind of growth along the steps or perpendicular to miscut direction. The saw-tooth steps may promote nanoparticle kind of growth on the triangular steps. Further, the step bunching is possible making the saw-tooth patterns more washed out. In Fig. 2 (a)-(e) the topographical patterns show traces of saw-tooth step flow growth. They are similar except the grain size becomes smaller with reducing thickness of the films.

The evolution of ferroelectric domains with variant film thickness, recorded through vertical PFM, is shown in Fig. 2 (f)-(j). The black and white contrasts in these images indicate polarization components pointing in down and up directions respectively. It has been documented that the BFO films have four and two polarization variants when deposited on exact and [100] vicinal substrates respectively.[18,26] The domains have large size and stripe patterns running perpendicular to the vicinal direction. Jang *et al* attributed such pattern formation to the relaxation of elastic-strain energy of the films on step surfaces without the necessity of two additional domain variants.[18] However, the vicinality along [110] direction yields saw-tooth step pattern, which may break the symmetry of stripe domains. It results in two variant domains with pattern intermediate of stripe and fractal. These domains have crystallographic orientation and straight wall features like stripe domains. However, the orientation is confined into smaller areas due to saw-tooth steps of the vicinal substrate. Such domains can have double advantages viz. better polarization due to two variants and better switchability resulting from step-edge dislocations.[20] Similar types of small bunches of striped ferroelastic domains are reported in compressively strained $TbMnO_3$ films deposited on $SrTiO_3$ substrates.[30]

An important feature of these domain patterns is the systematic reduction of domain size with the thickness. The contrast patterns with reducing dimensions were seen up to 10 nm thickness films. The 5 nm thickness film did not show the signature of domain formation probably due to crosstalk interference. On plain STO substrates, Daumont *et. al.*,[5] could not detect the contrast in 12 nm thickness BFO films whereas Catalan *et. al.*,[22] reported fractal domains up to 7 nm thickness films. They also reported the domain periodicity measured by Fourier analysis or by simply counting the number of domains across a straight segment. The average domain size departed from the classic LLK square root dependence on film thickness with the scaling exponent $\gamma = 0.59$. Alternatively, Zhao *et. al.*, proposed square dependence of domain period on thickness for $PbTiO_3$ ultrathin films.[31] The variation of domain width as a function of film thickness for the BFO films on vicinal substrates is shown in Fig. 3. The curve fitting to the law $w = Ad^{\gamma}$, where $w$ is the average domain width and d is the film thickness, gave the scaling exponent value $\gamma = 0.49 \pm 0.05$, which is very close to the LLK value of 0.5.[21] However, it is smaller than the value 0.59 reported by Catalan *et. al.*,[22] for the BFO thin films on plain substrate. This difference may be the attribute of substrate vicinality. On plain substrate the domains in thick film are larger with four polarization variants. The reduction of film thickness can cause rapid reduction in domain size making the scaling exponent greater than 0.5. On the other hand, use of [110] vicinal substrates yield small bunches of preferentially aligned domains. A low crystal anisotropy and pinning defects may be responsible for such appearance.[8] The terraces, steps, kinks, etc on [110] vicinal substrates can provide ample sites for the domain formations. Therefore, the domain size reduction with thickness may not be as rapid as in the case of plain substrates. In other words, [110] vicinal substrate assists growth of two variant, small size domains which follow LLK scaling closely. Such domains are more vulnerable to applied electric field and can be switched with lower field.[20]

There are quite a few riddles associated with the fundamental limit of ferroelectricity. In principle, the charge polarization occurs through the non-centrosymmetric arrangement of the atoms in perovskite unit cell. However, the celebrated concept of 'dead layers' in dielectric thin films indicate that there is a limit for the occurrence of polarization.[32,33] Stengel *et. al.*,[34] also proposed through the first-principle calculations an innovative concept of positive dead layer, which may actually enhance the ferroelectricity at the interface. The issue was complicate till the availability of reliable experimental technique to measure the polarization in ultrathin films, where high leakage current hampers the macroscopic measurements. The recent advancements in Scanning Probe Microscopy provided an effective tool to characterize localized ferroelectric response.[24,25] The electrical switching behavior has been reported in 1 nm $BaTiO_3$ [2], 30 nm $PbZrTiO_3$ [3] and 40 nm $BiFeO_3$ [35] films using PFM switching technique. The switching behavior of our films is shown in Fig. 4 (a)-(c). The characteristic butterfly loops were observed in amplitude signals of all the BFO films including the 5 nm thickness film (Fig. 4 a). Fig. 4 (b) shows the phase signal indicative of clear switching behavior. The collective piezo-response shown in Fig. 4 (c) also confirms that clear and complete polarization switching can be accomplished within the bias of ± 10 V. It gives an unambiguous evidence for the occurrence of ferroelectricity in BFO films as thin as 5 nm deposited on vicinal substrates. Although, Bea $et$ $al.,$[36] observed PFM pattern for 2nm thickness BFO films deposited on plain substrate, they have not reported switching behavior. Moreover, the use of vicinal substrate in the present study is more promising to obtain low-voltage switching.[20] However, it is difficult to quantify the magnitude of polarization and coercive field from SS-PFM data alone.[25] Nevertheless, the macroscopic measurements on thick films indicated that the effective switching of $BiFeO_3$ films on [110] vicinal substrates can be accomplished with lower voltage or reduced coercive field.[20] Therefore, the 5 nm $BiFeO_3$ films deposited on [110] vicinal substrate may satisfy the dimensional constraints for tunneling as well as the existence of switchable ferroelectricity. Recently, Maksymovych $et.$ $al.,$[3] have reported the simultaneous measurement of switchable polarization and local conductance in 30 nm tunnel barrier of PZT film.

## C. Electronic and transport properties

The transmission probability of electrons tunneling through an insulating barrier varies exponentially with the energy barrier height $\phi$. Therefore, understanding and controlling the $\phi$ value assumes central importance for applications concerning direct tunneling through ultra-thin insulators. For instance, the device resistance is directly related to this quantity in a magnetic tunnel junction. However for an insulator, $\phi$ is often assumed as equal to half the band gap ($E_g$) value of the material. The factor of half is a consequence of the definition of Fermi level placed halfway between the conduction band minimum (CBM) and valence band maximum. But this is true only if the electronic behavior is "free –electron" like. In a crystalline environment, due to band structure effects electrons propagate with an effective mass m*, which could be either greater or less than free electron mass, depending on the curvature of the energy bands. Therefore, the energy barrier height ($\phi$) is given by the relationship,

$$\varphi = \frac{E_g m^*}{2m} \qquad with \qquad m^* = \left(\frac{2\pi}{h}\right)^2 \left(\frac{d^2 E}{dk^2}\right) \qquad (1)$$

As explained above, $\phi$ equals half the band gap value only when m* = m and therefore, in general, assumes a different value. The most striking example is the case of crystalline MgO barrier, where the effect barrier height is measured to be 0.4 eV, almost a factor of 10 lower than half the band gap value (3.7 eV).[37, 38] It is well established by now that such low effective tunneling potentials observed in Fe/MgO/Fe are due to small effective masses at band edges of MgO. Below we investigate how effective masses contribute to the tunneling barrier height in the case of epitaxial high quality BFO films. We primarily focus on the rhombohedral (R3C symmetry) BFO structure. The tetragonal (P4mm) phase is also discussed briefly since it has been shown to have radically different structural characteristics compared to the rhombohedral phase[11, 12 ,16 ,17]. Both phases have potential for application in advanced spintronic devices with thin BFO film as part of a current carrying circuit. Thus, in addition to ferroelectric or magnetic behavior, electrical transport (tunneling) properties of this material are very important.

To account for above factors contributing to effective tunneling barrier on the level of accuracy beyond traditional Density Functional Theory (DFT), we employ highly accurate many body perturbation theory or Quasi-particle Self-Consistent GW (QSGW) theory.[39] More specifically we used scaled version[40] of the QSGW method, namely, we scale the difference of the self-energy and LDA exchange-correlation potential by a factor 0.8, $(\Sigma-V_{xc}) \rightarrow 0.8(\Sigma-V_{xc})$, that effectively takes into account renormalization of the polarization operator due to the electron-hole interaction. To quantify effective tunneling barrier or predict accurate relation for effective barrier of different phases of BFO, we need to focus on predictive theory of effective mass since band gap can be easily measured. In this context, one should be aware that band gaps within QSGW theory are somewhat overestimated due to electron-hole (exitonic) interaction effects, which are not included within the original GW theory formulation. Earlier, it has been shown that exitonic effects (additional ladder diagrams for polarization operator) account for most of the remaining errors in QSGW predicted band gaps.[41] Fortunately, it was shown that a simple scaling procedure with *universal* scaling factor 0.8 accounts for these exitonic effects and significantly improves QSGW results for band gaps.[40] Still, difference between theory and experiment for band gap within few decimals of electron-volt should be considered as indication of theory's good performance. Note that self-consistency enabled by QSGW method is expected to be important for calculating band gaps of transition metal based oxides.[39] Other important property contributing directly to effective tunneling barrier such as work function in case of transition metals and compounds also may require self-consistent treatment.[42] In Fig. 5 we compare electronic structure calculated within this many-body perturbation theory in comparison with earlier results obtained within the mean field type LDA/DFT or LDA+U theory [see Ref 43, for example]. The many-body nature of the QSGW method enables predictive calculations[39-42] of electronic and magnetic properties of materials with higher accuracy compared to LDA/DFT. The QSGW (solid) and LDA (dashed) electronic bands of R3c BFO are presented in Fig. 5a. Significant difference of the electronic bands calculated within the QSGW and the LDA should be noted as manifestation of important electron correlation effects in this material system. The conduction band minimum (CBM) falls at the same Z point in both methods, but the valence band maximum (VBM) is between the Z and Γ points in LDA while it is between F and Γ points in QSGW (there is also local maximum of QSGW valence bands between Z and L points that is very close to VBM). The effective masses are also different in LDA and QSGW as evident from the more dispersive QSGW bands. We calculated the effective mass at CBM (Z point) in direction of the BFO barrier growth. In the Z-F direction, $m_c^* = 2.2m$ in QSGW, while it is $m_c^* = 3.4m$ in LDA. The QSGW band gap of $E_g^{QSGW} = 3.3$ eV is in much better agreement with experimental BFO band gap $E_g^{expt} = 2.8$ eV. The LDA band gap value of about $E_g^{LDA} = 0.7$ eV, is in reasonable agreement with earlier reports of Neaton *et al.*[43] using the same technique. However, using screened exchange method Clark and Robertson[44] have reported a value of 2.8 eV. This result may seem as indication of better agreement with experiment but strictly speaking can not be interpreted this way since screened exchange method (as rightfully acknowledged by Clark and Robertson) is an approximation to GW theory.

For the estimation of the effective barrier height we employed simple model of Eq. (1). Using QSGW band gap $E_g^{QSGW}$ = 3.3 eV and effective mass at Z point (in Z-F direction) $m_c^*$ =2.2m, we obtain the QSGW effective barrier height as,

$$\varphi^{QSGW} = \frac{E_g m_c^*}{2m} = 3.63 eV$$

Here, we used the fact that QSGW effective mass at the top of the valence band at Z point is much larger than $m_c^*$ (QSGW band is essentially flat at Z). Eq. (1) gives LDA band gap $\phi^{LDA}$ = 1.19 eV for $E_g^{LDA}$ = 0.7 eV and $m_c^*$= 3.4m. We further investigated the electronic band structure of the BFO tetragonal phase as shown in Fig. 5 (b). Detailed comparison between BFO phases and with experiments will be published elsewhere. Even though this phase is strictly speaking monoclinic [11,12,16], the distortions are believed to be small (of the order of 1-2 degrees) and, in any case, the most striking feature of this phase is the extremely large c/a tetragonality factor close to 1.25. This is captured adequately in our calculations. We used the structural parameters for the hypothetical *P4mm* as reported in reference 13 and 45 which also enables consistent comparison between our QSGW and reported earlier LDA and LDA+U results. We found that the band structure is more dispersive compared to the R3c structure which reflects in very low electron mass (m* = 0.33m) at the G point in G-Z direction. The effective tunneling barrier height can be estimated according to the Eq. (1) as $\phi$= 0.38 eV for the QSGW band gap value of about 2.31 eV. Very recently there has been a report of a tetragonal BFO thin film with a band gap of 3.1 eV.[46] The dramatically low effective mass and, therefore, low barrier height of *P4mm* BFO is encouraging for spintronics applictions and we can predict based on this result that tunnel magnetoresistance devices with the nearly tetragonal BFO phase as the active barrier would have significantly lower resistance compared to *R3c* BFO. Therefore, favorable band structure "engineering" could be possible with strain stabilized tetragonal phase of BFO. Also note that the LDA bands predict a metallic state for the *P4mm* structure.

Now we report our preliminary investigation into the tunneling properties of rhombohedral 5 nm $BiFeO_3$ ultrathin film through local current vs voltage (I-V) measurements using conducting scanning probe tip of AFM. In Fig. 6 (a), we have plotted the current density (J) vs voltage (V) measured for the 5nm BFO film with voltage bias range of ±2V. Voltage was applied to the probe tip and current out of the sample was measured with the help of current amplifier (with a maximum gain of $10^9$). The non-linear behavior, characteristic of direct tunneling, was fitted to the Simmons phenomenological model used widely to estimate tunnel barrier properties.[47] The schematic diagram of the model implementation is shown in Fig. 6(b). Here, tunneling through a trapezoidal barrier is considered with similar metal electrodes on either side of the barrier, and the current density for intermediate voltage bias is given by,

$$J = \left( \frac{\gamma \sqrt{\varphi}}{d} \right) \exp\left( -A \sqrt{\varphi} \right) \left( V + \sigma V^3 \right) \tag{2}$$

where, $\gamma = e.\sqrt{2m} / 4b\pi^2 (h/2\pi)^2$, $A = 4\pi d \sqrt{2m} / h$, $\sigma = \frac{(A.e)^2}{96.\phi.d^2} - \frac{A.e^2}{32.d.\phi^{\frac{3}{2}}}$

b = 23/24, m = free electronic mass = $9.31 \times 10^{-31}$ kg, h = Planck's constant = $6.63 \times 10^{-34}$ J.s, e = electronic charge = $1.603 \times 10^{-19}$ C, d is the effective barrier thickness, and $\varphi$ is the effective mean barrier height.

As equation 2 shows, tunneling is linear in the low voltage bias regime and non-linearity is captured in the cubic term at higher voltages. The measured current density for the 5 nm film is extremely low and implies extremely high resistance-area (RA) product factor for BFO based spintronic devices. Such high resistance values are consistent with existing reports.[48] An asymmetry observed in the experimental curve and deviation of the curve from the Simmons fit can be attributed to dissimilar electrodes in the experimental setup. With a tip radius of ~10 nm, and barrier width of 5nm we obtain an effective barrier height of 4.15 eV for positive voltage values and for m*=m. This is in favorable agreement with the QSGW barrier height estimation of 3.63 eV. Deviation is significant if we compare the LDA barrier height value $\phi$=1.19 eV. However, interestingly, if we use the experimental band gap value of 2.8 eV to estimate $\phi$ with get $\phi^{LDA}$= 4.76 eV where as $\phi^{QSGW}$ = 3.08eV for their respective effective mass values. The value from Simmons fit is halfway between these numbers. It is to be noted that there are other ways to fit the experimental curves within the tunneling regime. For instance, Brinkman model[49] can explicitly include barrier asymmetry ($\Delta\phi$). Within this framework, the results were $\phi$= 2.2 eV, $\Delta\phi$ = 5.9 eV giving an average barrier height of ($\phi + \Delta\phi/2$) equals 5.07eV, which is larger than the Simmons fit value. The barrier thickness from the Brinkman model fit was found to be 1.4 nm. The unexpectedly large barrier asymmetry value and the underestimated barrier thickness leads us to conclude that the Brinkman fit, in this case, is less physical than Simmons. An alternate, and probably the best method, of estimating the experimental barrier height is from the WKB approximation, where the resistance-area product of the MTJ incorporating the BFO barrier is measured as a function of the barrier thickness and the average barrier height is extracted as fitting result. This at present is beyond the scope of this work as it involves fabrication of tunnel junction devices with well defined sizes. In the experience of the authors the Simmons fit value is usually quite close to the WKB estimation and therefore was used in this work.

# IV CONCLUSION

We deposited $SrRuO_3$ bottom layered 5-100 nm thickness $BiFeO_3$ films on the vicinal $SrTiO_3$ substrates with $4^0$ miscut along [110] direction. The choice of substrate was to yield two variant domain structures with better ferroelectric properties. The epitaxial and constrained films showed traces of saw-tooth step flow growth in topographical features. The Piezo-response Force Microscopy revealed two variant, preferentially aligned domains. The domain width scaling with film thickness closely followed the KKL law. A clear switching was observed through Switching Spectroscopy Piezo-response Force Microscopy in all the films including the ultrathin 5 nm film. From local I-V measurements, we conclude BFO to be a high resistance barrier material with an effective barrier potential of over 4 eV.

We have also performed beyond local density functional *ab-initio* calculations of the band structure of BFO using QSGW methods, backed up by nanoscale transport measurements. We have, in particular, focused in estimating the effective barrier height for tunnel electrons. We find that a large effective mass at CBM leads to a high effective barrier height (3.6eV) for the R3c phase which is confirmed through local current-voltage tunneling measurements using the conducting AFM technique and Simmons model for direct tunneling. We also employed QSGW theory for tetragonal *P4mm* phase of BFO and found a significantly lower effective mass and effective tunneling barrier (0.38 eV) assuming defect free film, band alignment in the center of the band gap and AF (001) magnetic structure. We anticipate that the nearly-tetragonal phase with slight monoclinic distortions to show similar properties making this phase much more attractive for spintronics application.

**ACKNOWLEDGEMENT**

This work was supported by ONR under Grant No. N00014-09-1-0119 and NSF NIRT under Grant No. CMS-0609377. A portion of this research was conducted at the Center for Nanophase Materials Sciences, which is sponsored at Oak Ridge National Laboratory by the Division of Scientific User Facilities, U.S. Department of Energy.  OM and SF acknowledge the CNMS User computer support by Oak Ridge National Laboratory Division of Scientific User facilities, Office of Basic Energy Sciences, U.S. Department of Energy.

**CAPTION OF FIGURES:**

**Fig 1** (a) The X-ray diffraction patterns for 100 (V119), 50 (V120), 20 (V121) and 10 (V122) nm $BiFeO_3$ films deposited with $SrRuO_3$ bottom layer on [110] vicinal $SrTiO_3$ substrates. Inset: Resolved (*001*) peaks.

Representative omega (b) and phi (c) scans to reveal crystalline and epitaxial nature.

**Fig 2** The surface topography scans (a)-(e) and corresponding ferroelectric domain structures (f)-(j) of 2 μm x 2 μm area of 100, 50, 20, 10 and 5 nm thickness $BiFeO_3$ films respectively.

**Fig 3** The variation of domain width as a function of film thickness for BFO films deposited on [110] vicinal substrates.

**Fig 4** Switching behavior of BFO films of variable thickness as revealed through SS-PFM: (a) amplitude, (b) phase, and (c) piezo-response signals as a function of tip bias.

**Fig 5** The Electronic band structure of a) rhombohedral (R3c) and b) tetragonal (P4mm) phases of $BiFeO_3$ calculated along the high symmetry lines in Brillouin zone within the scaled QSGW (see text) (solid lines) and DFT-LDA (dash-dotted lines) theory. The F-Z line corresponds to measured direction of transport for the R3c phase.

**Fig 6** (a) Experimental J-V curve (black) and Simmons fit (red). The mean barrier height obtained is 4.15 eV (b) Simmons barrier model for 5nm $BiFeO_3$ film sandwiched between bottom SRO and Ir-Pt top electrode.

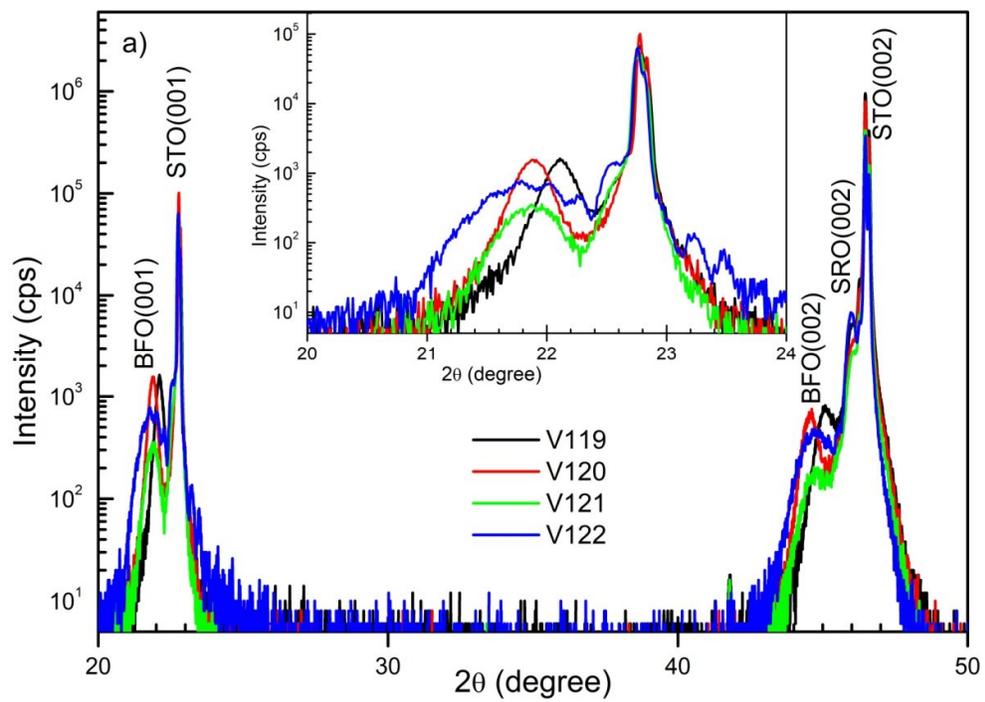

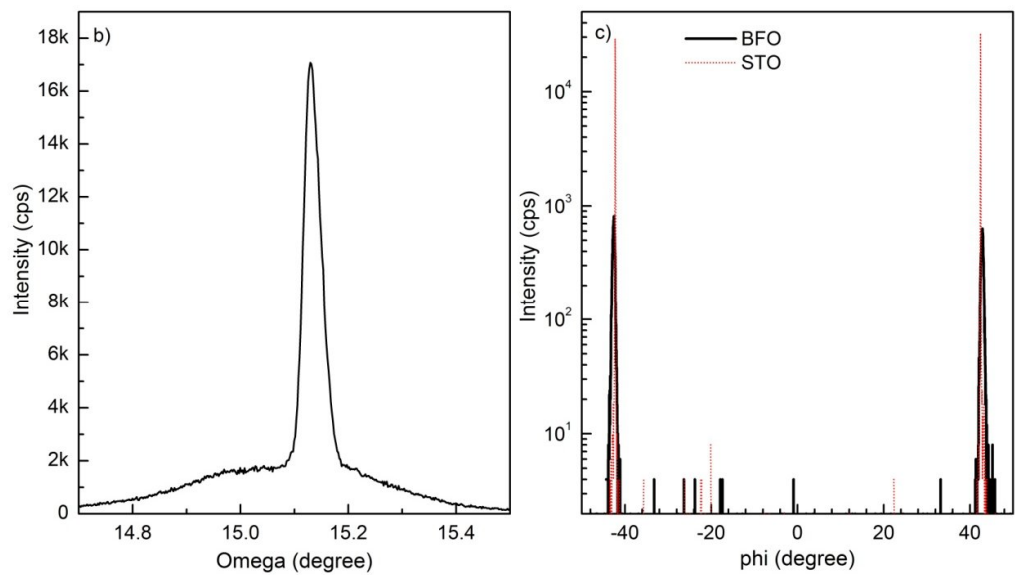

Fig. 1

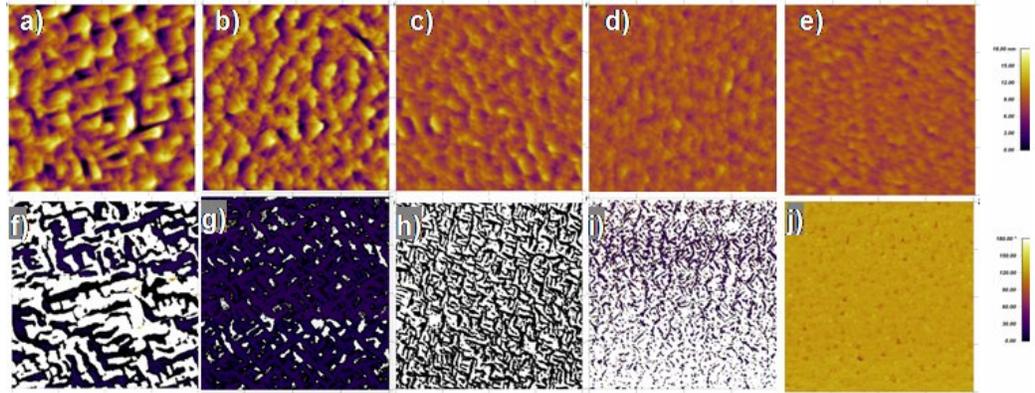

Fig 2

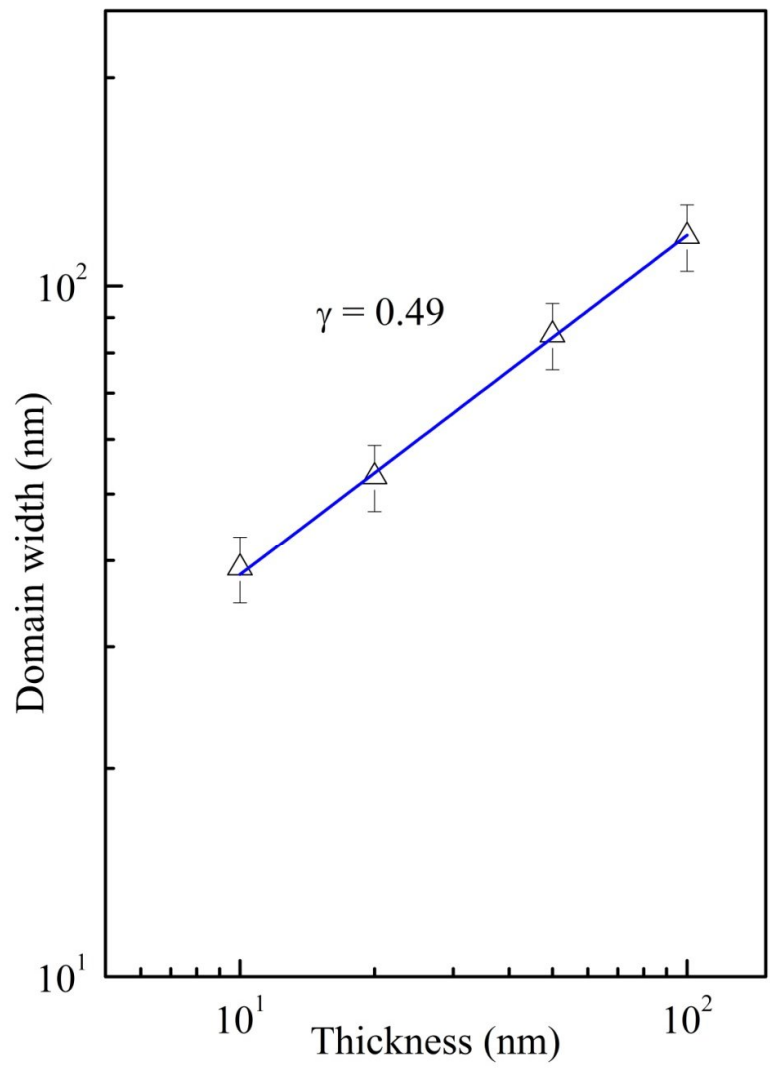

Fig 3

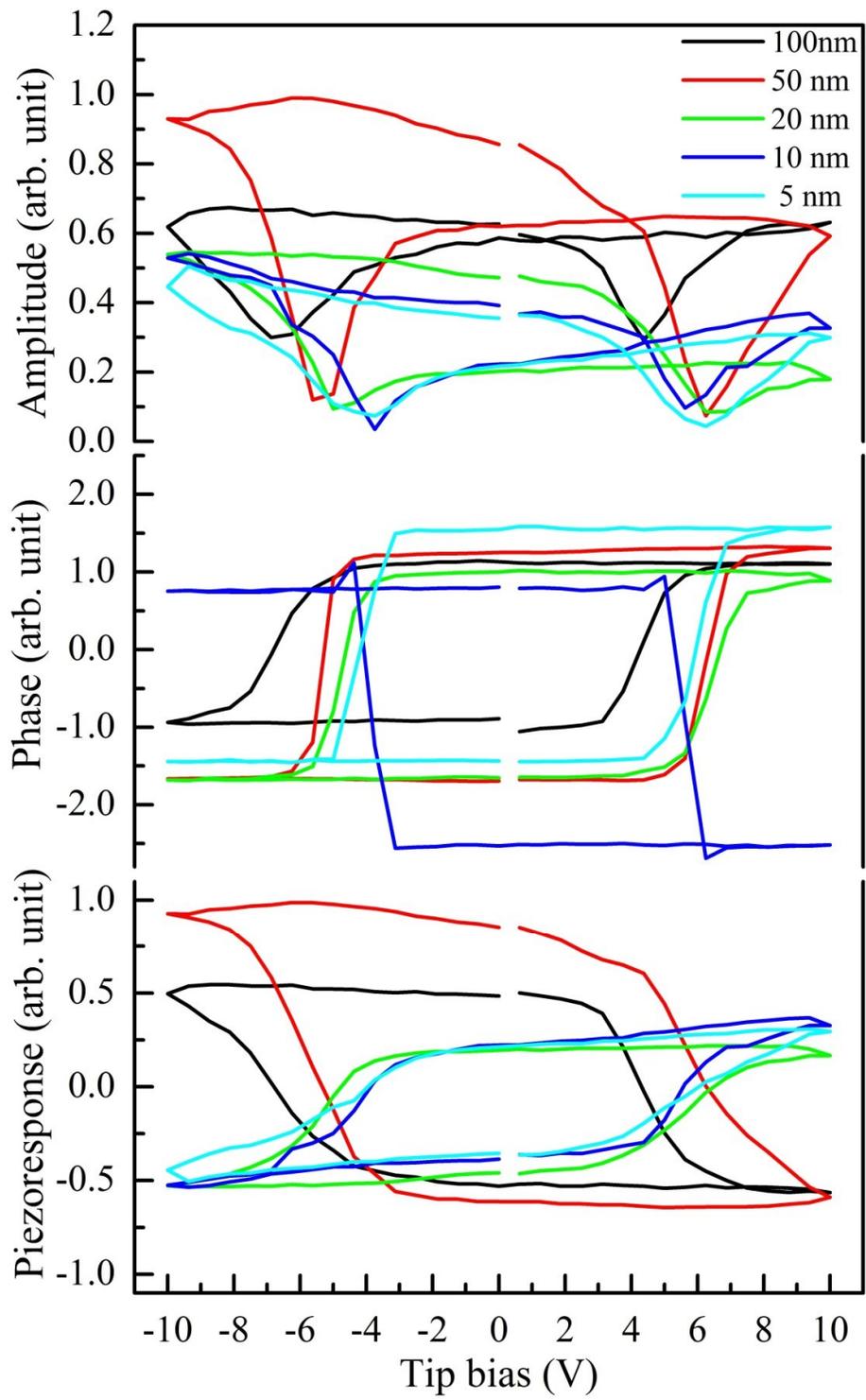

Fig 4

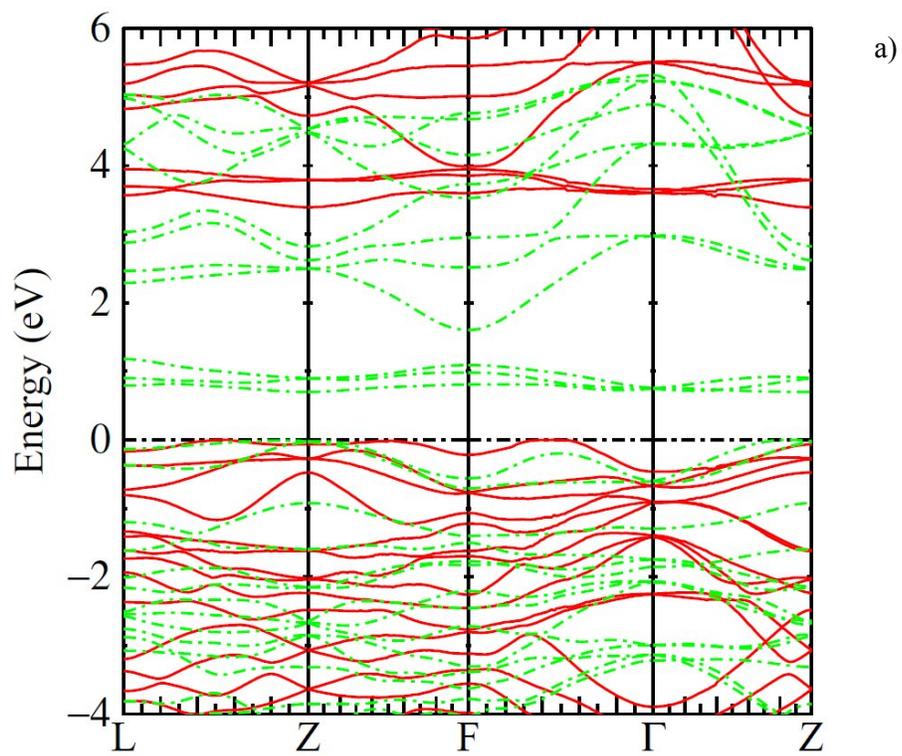

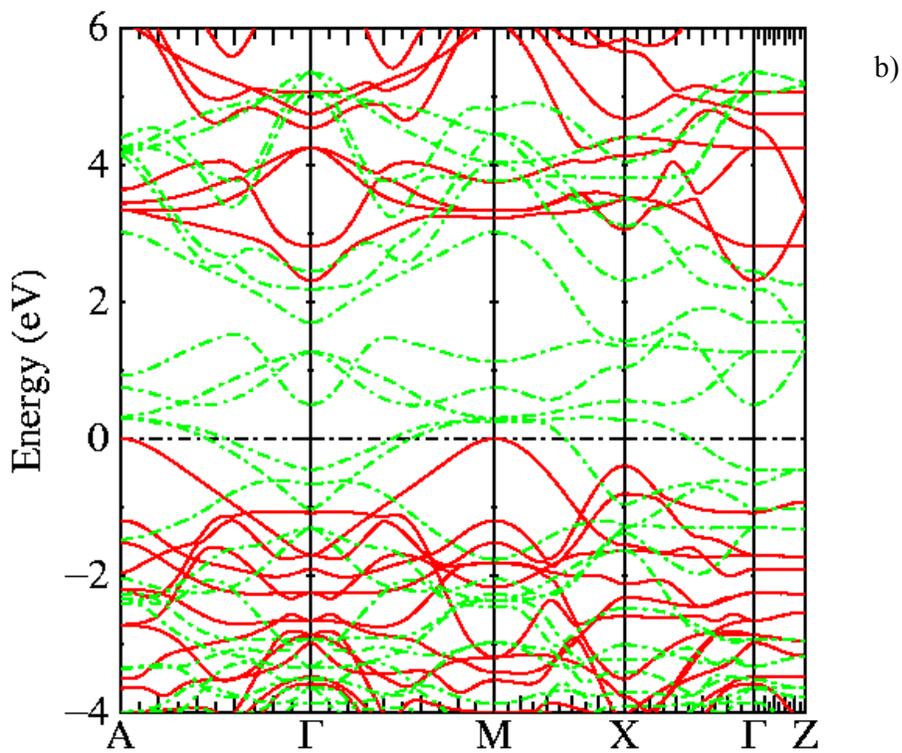

Fig 5

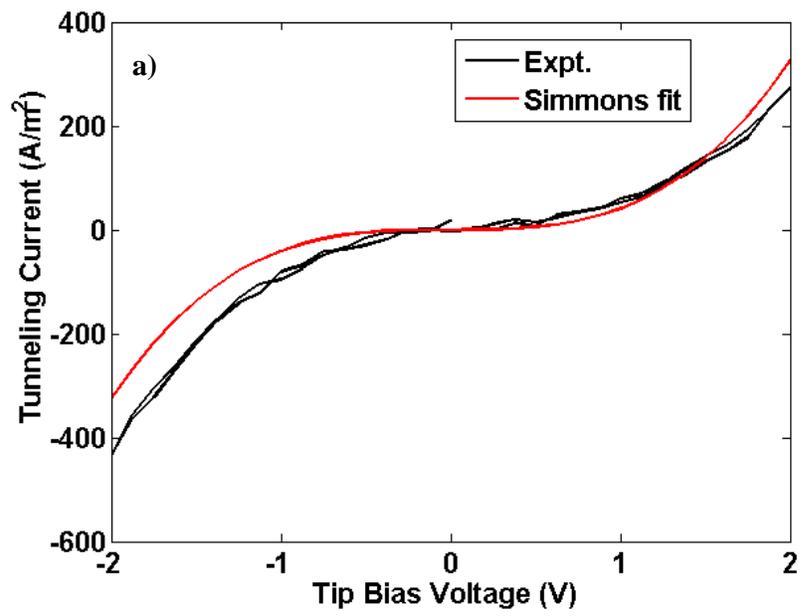

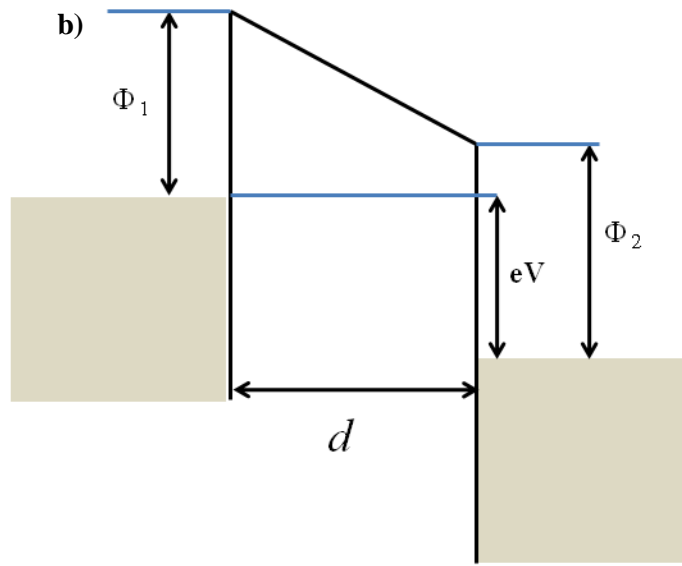

Fig 6